%% file: mmu2023_perotto.tex
\documentclass{webofc}
\usepackage[varg]{txfonts}   
\usepackage{xcolor}
\definecolor{bleu}{RGB}{30,144,255}
\definecolor{bleu2}{RGB}{26,102,255}
\newcommand*\comment[1]{#1}

\begin{document}
\title{The NIKA2 Sunyaev-Zeldovich Large Program}
%
\subtitle{Sample and upcoming product public release}

\author{%
  \lastname{L.~Perotto}\inst{\ref{LPSC}}\fnsep\thanks{ laurence.perotto@lpsc.in2p3.fr}
  \and  R.~Adam \inst{\ref{OCA}}
  \and  P.~Ade \inst{\ref{Cardiff}}
  \and  H.~Ajeddig \inst{\ref{CEA}}
  \and  P.~Andr\'e \inst{\ref{CEA}}
  \and  E.~Artis \inst{\ref{Garching}}
  \and  H.~Aussel \inst{\ref{CEA}}
  \and  R.~Barrena \inst{\ref{Tenerife1}, \ref{Tenerife2}}
  \and  I.~Bartalucci \inst{\ref{Milano}}
  \and  A.~Beelen \inst{\ref{LAM}}
  \and  A.~Beno\^it \inst{\ref{Neel}}
  \and  S.~Berta \inst{\ref{IRAMF}}
  \and  L.~Bing \inst{\ref{LAM}}
  \and  O.~Bourrion \inst{\ref{LPSC}}
  \and  M.~Calvo \inst{\ref{Neel}}
  \and  A.~Catalano \inst{\ref{LPSC}}
  \and  M.~De~Petris \inst{\ref{Roma}}
  \and  F.-X.~D\'esert \inst{\ref{IPAG}}
  \and  S.~Doyle \inst{\ref{Cardiff}}
  \and  E.~F.~C.~Driessen \inst{\ref{IRAMF}}
  \and  G.~Ejlali \inst{\ref{Tehran}}
  \and  A.~Ferragamo \inst{\ref{Roma}}
  \and  A.~Gomez \inst{\ref{CAB}} 
  \and  J.~Goupy \inst{\ref{Neel}}
  \and  C.~Hanser \inst{\ref{LPSC}}
  \and  S.~Katsioli \inst{\ref{Athens_obs}, \ref{Athens_univ}}
  \and  F.~K\'eruzor\'e \inst{\ref{Argonne}}
  \and  C.~Kramer \inst{\ref{IRAMF}}
  \and  B.~Ladjelate \inst{\ref{IRAME}} 
  \and  G.~Lagache \inst{\ref{LAM}}
  \and  S.~Leclercq \inst{\ref{IRAMF}}
  \and  J.-F.~Lestrade \inst{\ref{LERMA}}
  \and  J.~F.~Mac\'ias-P\'erez \inst{\ref{LPSC}}
  \and  S.~C.~Madden \inst{\ref{CEA}}
  \and  A.~Maury \inst{\ref{CEA}}
  \and  P.~Mauskopf \inst{\ref{Cardiff},\ref{Arizona}}
  \and  F.~Mayet \inst{\ref{LPSC}}
  \and  A.~Monfardini \inst{\ref{Neel}}
  \and  A.~Moyer-Anin \inst{\ref{LPSC}}
  \and  M.~Mu\~noz-Echeverr\'ia \inst{\ref{LPSC}}
  \and  A.~Paliwal \inst{\ref{Roma}}
  \and  G.~Pisano \inst{\ref{Roma}}
  \and  E.~Pointecouteau \inst{\ref{IRAP}}
  \and  N.~Ponthieu \inst{\ref{IPAG}}
  \and  G.~W.~Pratt \inst{\ref{CEA}}
  \and  V.~Rev\'eret \inst{\ref{CEA}}
  \and  A.~J.~Rigby \inst{\ref{Leeds}}
  \and  A.~Ritacco \inst{\ref{INAF}, \ref{ENS}}
  \and  C.~Romero \inst{\ref{Pennsylvanie}}
  \and  H.~Roussel \inst{\ref{IAP}}
  \and  F.~Ruppin \inst{\ref{IP2I}}
  \and  K.~Schuster \inst{\ref{IRAMF}}
  \and  A.~Sievers \inst{\ref{IRAME}}
  \and  C.~Tucker \inst{\ref{Cardiff}}
  \and  G.~Yepes \inst{\ref{Madrid}}
}
\institute{
  Universit\'e Grenoble Alpes, CNRS, Grenoble INP, LPSC-IN2P3, 38000 Grenoble, France
  \label{LPSC}
  \and
  Universit\'e C\^ote d'Azur, Observatoire de la C\^ote d'Azur, CNRS, Laboratoire Lagrange, France 
  \label{OCA}
  \and
  School of Physics and Astronomy, Cardiff University, CF24 3AA, UK 
  \label{Cardiff}
  \and
  Universit\'e Paris-Saclay, Université Paris Cité, CEA, CNRS, AIM, 91191, Gif-sur-Yvette, France
  \label{CEA}
  \and	
  Max Planck Institute for Extraterrestrial Physics, 85748 Garching, Germany
  \label{Garching}
  \and
  Instituto de Astrof\'isica de Canarias, E-38205 La Laguna, Tenerife, Spain 
  \label{Tenerife1}
  \and
  Univ. de La Laguna, Departamento de Astrofísica, E-38206 La Laguna, Tenerife, Spain 
  \label{Tenerife2}
  \and
  INAF, IASF-Milano, Via A. Corti 12, 20133 Milano, Italy
  \label{Milano}
  \and
  Aix Marseille Univ, CNRS, CNES, LAM, Marseille, France
  \label{LAM}
  \and
  Universit\'e Grenoble Alpes, CNRS, Institut N\'eel, France
  \label{Neel}
  \and
  Institut de Radioastronomie Millim\'etrique (IRAM), 38406 Saint Martin d’H\`eres, France
  \label{IRAMF}
  \and 
  Dipartimento di Fisica, Sapienza Universit\`a di Roma, I-00185 Roma, Italy
  \label{Roma}
  \and
  Univ. Grenoble Alpes, CNRS, IPAG, 38000 Grenoble, France 
  \label{IPAG}
  \and
  Institute for Research in Fundamental Sciences (IPM), Larak Garden, 19395-5531 Tehran, Iran
  \label{Tehran}
  \and
  Centro de Astrobiolog\'ia (CSIC-INTA), Torrej\'on de Ardoz, 28850 Madrid, Spain
  \label{CAB}
  \and
  National Observatory of Athens, IAASARS, GR-15236, Athens, Greece
  \label{Athens_obs}
  \and
  Faculty of Physics, University of Athens, GR-15784 Zografos, Athens, Greece
  \label{Athens_univ}
  \and
  High Energy Physics Division, Argonne National Laboratory, Lemont, IL 60439, USA
  \label{Argonne}
  \and  
  Instituto de Radioastronom\'ia Milim\'etrica (IRAM), E~18012 Granada, Spain
  \label{IRAME}
  \and
  LERMA, Observatoire de Paris, PSL Research Univ., CNRS, Sorbonne Univ., UPMC, 75014 Paris, France  
  \label{LERMA}
  \and
  School of Earth \& Space and Department of Physics, Arizona State University, AZ 85287, USA
  \label{Arizona}
  \and
  Universit\'e de Toulouse, UPS-OMP, CNRS, IRAP, 31028 Toulouse, France
  \label{IRAP}
  \and
  School of Physics and Astronomy, University of Leeds, Leeds LS2 9JT, UK
  \label{Leeds}
  \and
  INAF-Osservatorio Astronomico di Cagliari, 09047 Selargius, Italy
  \label{INAF}
  \and   
  LPENS, ENS, PSL Research Univ., CNRS, Sorbonne Univ., Universit\'e de Paris, 75005 Paris, France 
  \label{ENS}
  \and
  Department of Physics and Astronomy, University of Pennsylvania, PA 19104, USA
  \label{Pennsylvanie}
  \and
  Institut d'Astrophysique de Paris, CNRS (UMR7095), 75014 Paris, France
  \label{IAP}
  \and
  University of Lyon, UCB Lyon 1, CNRS/IN2P3, IP2I, 69622 Villeurbanne, France
  \label{IP2I}
  \and
  Departamento de F\'isica Te\'orica and CIAFF, Facultad de Ciencias, Universidad Aut\'onoma de Madrid, 28049 Madrid, Spain
  \label{Madrid}
}

\abstract{%
  The NIKA2 camera operating at the IRAM 30-m telescope excels in 
  high-angular resolution mapping of the thermal Sunyaev-Zel’dovich effect towards galaxy clusters at intermediate and high-redshift. As part of the NIKA2 guaranteed-time, the SZ Large Program (LPSZ) aims at tSZ-mapping a representative sample of SZ-selected galaxy clusters in the catalogues of the Planck satellite and of the Atacama Cosmology Telescope, and also observed in X-ray with \emph{XMM-Newton} or \emph{Chandra}. Having completed observations in January 2023, we present tSZ maps of 38 clusters spanning the targeted mass ($3 < M_{500}/10^{14} M_{\odot} < 10$) and redshift ($0.5 < z < 0.9$) range. The first in-depth studies of individual clusters highlight the potential of combining tSZ and X-ray observations at similar angular resolution for accurate mass measurements. These were milestones for the development of a standard data analysis pipeline to go from NIKA2 raw data to the thermodynamic properties of galaxy clusters for the upcoming LPSZ data release. Final products will include unprecedented measurements of the mean pressure profile and mass-observable scaling relation using a distinctive SZ-selected sample, which will be key for ultimately improving the accuracy of cluster-based cosmology.}
\maketitle
\section{Introduction}
\label{Intro}
 
Obtaining accurate measurements of the mass of galaxy clusters and understanding deviations from the self-similar model due to baryonic physics are key challenges when using these valuable tracers of large scale structures as cosmological probes~\cite{Pratt2019}. High-angular resolution observation of galaxy clusters through the thermal Sunyaev-Zel'dovich (tSZ) effect~\cite{SZ} presents a promising \comment{way forward}. The tSZ effect is independent of redshift and yields the Compton parameter reflecting electronic pressure \comment{within the intra-cluster medium}. High-angular resolution mapping the tSZ effect towards clusters provides a radial pressure profile estimate via straightforward deprojection. When combined with X-ray observations allowing density profile deprojection~\cite{Bartalucci2017}, cluster masses can be inferred assuming sphericity and hydrostatic equilibrium~\cite{Pratt2019}.

Such measurements require advanced millimetre-domain experiments that offer a sufficiently large field of view to cover the typical sizes of clusters and high angular resolution to resolve their detailed structure.  NIKA2, at the IRAM 30-meter telescope, uniquely combines a wide 6.5' \comment{field-of-view diameter} with 17.6 and 11.1 arcsecond FWHM Gaussian beam at 150 and 260~GHz, respectively~\cite{NIKA2-performance}.

The SZ Large Program of NIKA2 (LPSZ; 300 hours of guaranteed time; P.I. F.~Mayet \& L.~Perotto) maps the tSZ effect at high resolution for a representative sample of 38 galaxy clusters selected from the Planck and ACT catalogues and covering a wide range of mass and redshift. This extends existing calibration samples (e.g. \cite{Pratt2009}, \cite{Planck2013}) to higher redshifts or lower masses. Using this unique sample, our main goal is to improve the measurement of the mean pressure profile of clusters and the scaling relation between SZ observable and mass. These key tools for many SZ cosmological analyses will ultimately contribute to improving the accuracy of cluster-based cosmology.

\section{The LPSZ cluster sample}
\label{Sect1}

The selection of the LPSZ sample occurred in 2015, utilizing the available Planck~\cite{PSZ1} and ACT~\cite{ACT2015} catalogues at that time. Two primary objectives guided this selection: to ensure \comment{representativity with respect to morphology} and to achieve homogeneous coverage across the mass-redshift plane, including redshifts greater than $0.5$ and a mass range as broad as permitted by the total 300 hours of NIKA2 Guaranteed Time of observation.
To this end, clusters were selected based on their detected SZ signal amplitude, effectively minimizing biases linked to cluster morphology or dynamical state. For an homogeneous distribution across the mass-redshift plane, 10 mass-redshift bins were defined by partitioning the mass range from $3\times 10^{14}$ to $10^{15} M_{\odot}$ into 5 logarithmic spaced intervals, and \comment{the redshift range from 0.5 to 0.9 in two linear bins.} In each mass-redshift bin, 5 clusters were randomly selected using their integrated Compton parameter measurements and the mass-observable relation from \cite{A10}.  Consequently, the original 2015 selection comprised 45 clusters, 35 from the Planck catalogue, and 10 ACT clusters covering lower mass ranges. The choice of the number of bins and clusters per bins were optimized from observation time consideration. For each cluster, the target \comment{was to obtain} a $3 \sigma$ measurement of the Compton parameter radial profile at a radius $\theta_{500}$, the angular diameter subtended by the typical cluster size. This size is given by $R_{500}$, the radius of a sphere centered on the cluster in which the average density is 500 times the critical density of the Universe.

Observations were conducted from October 2017 to January 2023 during 29 observation campaigns, each lasting one to two weeks, spread between October and March annually. Each campaign was guided by a detailed observation plan, with targets selected based on 1/ visibility, 2/ mass and redshift to gradually fill the LPSZ mass-redshift bins, and 3/ any additional decisive information from Planck cluster catalogue follow-up \comment{programmes}. Notably, 4 clusters from the \comment{initial Planck selection} were later either identified as false detections or appeared very faint in X-ray follow-ups. 
Additionally, 2 clusters lacked sufficient X-ray emissions for the necessary density profile deprojection to estimate hydrostatic mass. The former 4 objects were discarded, and the latter 2 received lower observation priority and ultimately remained unobserved. Following target selection, allocated observation times were re-evaluated iteratively, leveraging data from prior campaigns to meet the LPSZ SNR criterion. While this in-flight re-evaluation ensured each observed cluster supported LPSZ main goals, it also extended total observation time. Consequently, 4 ACT clusters, already at the visibility limit of the IRAM 30-meter telescope during observation months, were excluded in favour of redistributing observation time to Planck clusters. \comment{Additionally}, 3 clusters, replacing the false detection within the corresponding mass-redshift boxes, were selected from the latest Planck and ACT catalogues~\cite{PSZ2, ACT2018}.

\begin{figure}[!h]
\centering
\includegraphics[width=1.1\textwidth]{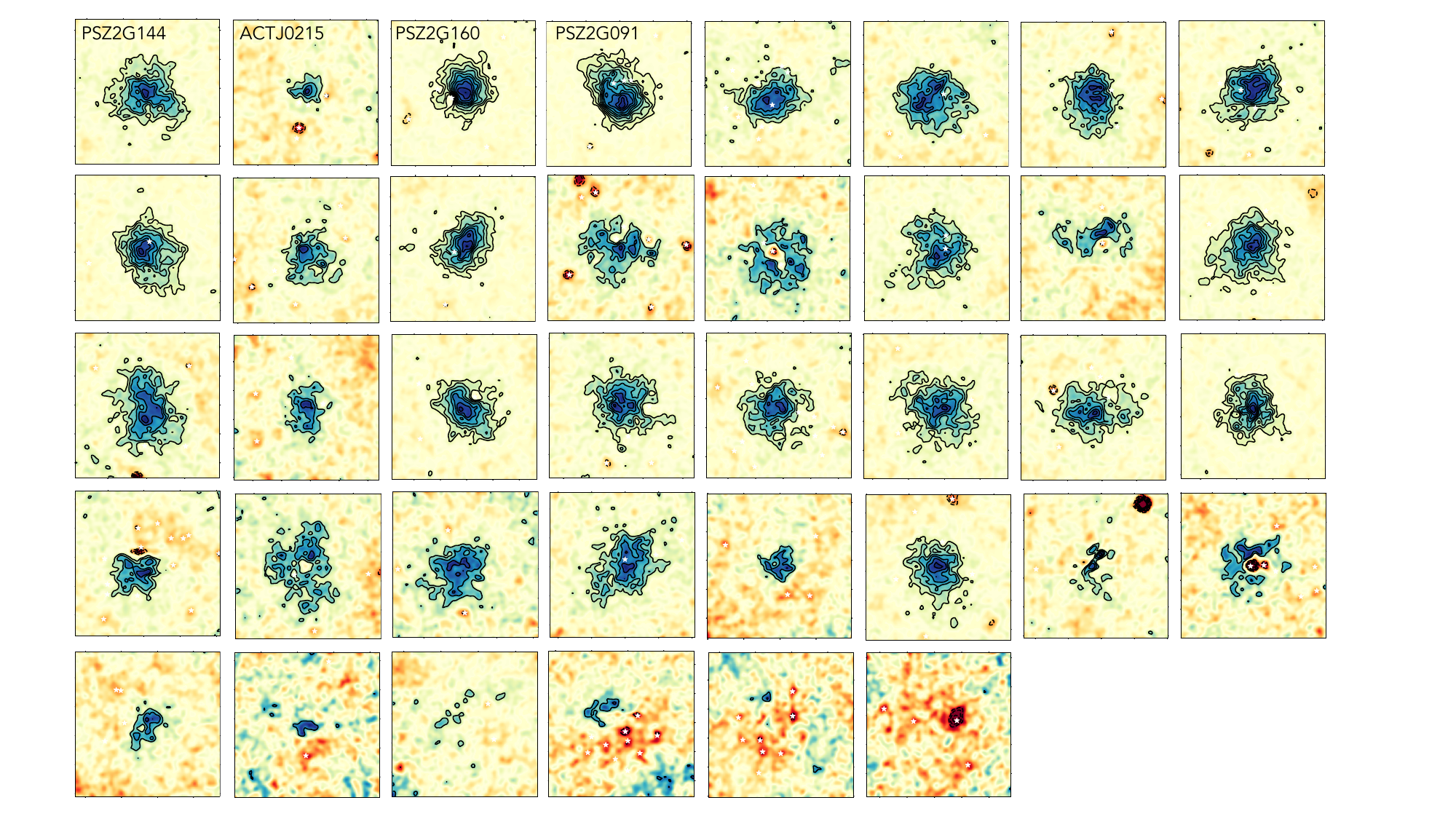}
\caption{Preliminary analysis of the LPSZ sample after the end of observation: 
  150~GHz surface brightness maps of the 38 observed clusters, colour-coded arbitrarily. 
  The top four maps correspond to the first in-depth analysis of the following LPSZ clusters: $\rm{PSZ}2\, G144.83+25.11$~\cite{Ruppin_PSZ2G144}, ACT-CL J$0215.4+0030$~\cite{Keruzore_ACTJ0215}, PSZ2 G$160.83+81.66$ (a.k.a. CLJ$1226.9+3332$)~\cite{Miren_CLJ1227} and PSZ2 G$091.83+26.11$~\cite{EA_proceeding}, respectively. Using the evaluation of the SNR levels (showed as black contours) and prior information from the Planck and ACT catalogues, we find that 30 clusters are conservatively mapped at the requested SNR level for LPSZ core science goals. \comment{This sample is showed by the maps of the 4 aforementioned clusters and the randomized 26 remaining clusters.}
  The five next maps show detected clusters for which refined analyses are needed to enhance the map quality.
The three last maps show highly-contaminated or noise-dominated cases.}
\label{mosaic}       
\end{figure}

In conclusion, 38 galaxy clusters were observed. Figure~\ref{mosaic} presents the 150~GHz map for each cluster, obtained during preliminary data quality assessment. These maps are not the final versions for publication. The iterative observation process facilitated a uniform coverage of the mass-redshift plane defined by LPSZ, ensuring a minimum of 3 clusters per mass-redshift box. Among these, the 150~GHz map for 30 clusters already meets the target SNR. In 5 other cases, the current SNR of the map should be effectively improved in resorting to a refined version of the analysis. 
Only 3 clusters exhibit a preliminary map that casts doubt on the feasability of deprojecting a resolved pressure profile. 
Detailed study of four LPSZ cluster has been published, as will be discussed in the subsequent section.

\section{Recent results}
\label{Sect2}

Parallel to the observation and data quality analysis, the first in-depth studies of individual clusters have been published. Following the publication of the first galaxy cluster mapped through the tSZ effect with NIKA2~\cite{Ruppin_PSZ2G144}, we investigated  a challenging low-mass, high-redshift cluster, demonstrating the feasibility of mass estimation across the entire sample~\cite{Keruzore_ACTJ0215}. These initial studies have marked important milestones for estimating systematic effects and preparing cosmological results. 
Furthermore, the method described in \cite{Keruzore_ACTJ0215} led to panco2, a public code for pressure profile estimation from SZ maps obtained from any millimeter-wave instrument, as detailed in \cite{Keruzore_PANCO2}.

In \cite{Miren_CLJ1227}, we found the dominant systematic uncertainty on mass estimation to be related to the choice of the cluster pressure profile model, rather than residual noise or large-scale angular filtering induced by data reduction. Studies combining LPSZ observations with gravitational lensing reconstructions in the CLASH \comment{programme}~\citep{Ferragamo_lensing, Miren_CLJ1227} have demonstrated the potential of this approach to study the hydrostatic bias, linking the mass estimated under the assumption of hydrostatic equilibrium with the total mass of clusters as probed via gravitational lensing.

Additionally, leveraging high-resolution X-ray and tSZ observations, we studied a highly disturbed, massive, high-redshift cluster~\cite{EA_proceeding}. This provided an extreme case for investigating morphology and dynamical state impacts on mass estimates. This system being consistent with a scenario of a major merger event involving two main halos, we identified map regions most compatible with the hydrostatic equilibrium, facilitating the hydrostatic mass estimation of both halos.    

Lastly, to assess LPSZ sample sensitivity to cluster physics and deviations from underlying assumptions in cluster-based cosmology, we constructed a series of LPSZ twin samples selected from the state-of-the-art simulation of \emph{The Three Hundred} project~\cite{Cui2018, Paliwal2022, Miren_twin}.

\section{Standard analysis and upcoming data release}
\label{Sect3}

We \comment{developed} an analysis pipeline to go from Time-Ordered Information (TOI) to the products of interest for cosmology. This pipeline is divided in three blocks.

The first stage consists in going from the raw TOI to surface brightness maps, whose noise properties and angular scale filtering are well characterized. To that aim, we use the NIKA2 collaboration IDL data reduction pipeline, initially developed for the performance assessment~\cite{NIKA2-performance}, with modifications to optimally preserve the signal at large angular scales while subtracting the correlated noise. The remaining scale filtering induced by this process is modelled with a transfer function estimated from processing simulated inputs through the pipeline. The noise properties of the maps are characterized both with the angular power spectrum of sign-flipped co-added maps, and with sample count maps. 
Finally, we leverage the 260-GHz map, in which residual noise is expected to dominate over the SZ positive signal, for the detection of point sources in the cluster neighbourhood. Point source photometry using the 260~GHz map allows us to set an upper limit of the contamination from sub-mm sources to the tSZ signal in the 150~GHz map. In addition, we systematically \comment{search for} point sources detected around LPSZ clusters in ancillary radio, submillimetric and NIR point source catalogues.

The second stage is the full characterization of the thermodynamic properties of each cluster. We use panco2 for the deprojection of 3D spherical pressure profiles from NIKA2 maps. This method resorts to forward modelling of the NIKA2 map at 150~GHz, featuring the tSZ signal, realistic correlated noise and filtering, and the contribution of detected point sources, to estimate a spherical pressure profile. As in previous LPSZ work, we combine \comment{the pressure profile obtained from NIKA2 data} with the density profile reconstructed from \emph{XMM-Newton} observation using the method described in~\cite{Bartalucci2017}, for the full characterization of the thermodynamic properties of the clusters. In particular, we infer the \comment{radial hydrostatic mass profile}, from which we estimate $M_{500}$, the mass enclosed in a sphere of radius $R_{500}$.

The last stage is the sample level analysis to build the products of interest for cluster-based cosmology. Using a simulation suite featuring realistic NIKA2 noise, we validate our methodology for estimating the mass-observable scaling relation~\cite{Alice}, as well as for deriving the mean pressure profile~\cite{Corentin}.

We expect to publicly release the main LPSZ products no later than mid-2025. This release will encompass
several components : i) data products, such as frequency maps, noise characterization,
and filtering, ensuring reproducibility of our results and facilitating subsequent
combined studies; ii) 
a comprehensive characterization of clusters in the LPSZ sample, including their thermodynamic
properties, and their hydrostatic masses; iii) derived products of interest, such as a catalogue
of point sources detected in LPSZ maps, some of which may correspond to high-redshift galaxies magnified by the lensing of clusters; and finally iv) final products of interest for cosmology, particularly the mean pressure profile and the mass-observable relation.

\section{Conclusion}
\label{Conclu}

Within the framework of NIKA2 LPSZ, a 300-hour guaranteed-time \comment{programme}, we have achieved resolved SZ mapping of a SZ-selected sample of 38 clusters. The sample spans a redshift range from $0.5$ to $0.9$ and covers an order of magnitude in mass.

First in-depth analyses of four clusters showcase the potential of resolved SZ observations to open up novel insights on cluster physics and to obtain accurate mass estimates. In particular, new perspectives were explored on the physics of disturbed clusters or on hydrostatic-to-lensing mass bias.

Having concluded the observations in January 2023, our preliminary assessments indicate that 35 clusters have been successfully mapped with the expected signal-to-noise ratio. We have a robust foundation for fulfilling the primary objectives of the LPSZ. In preparation \comment{for} the forthcoming public data release, we are refining a standard analysis pipeline to go from raw TOI to the products of interest for Cosmology. These final products will include unprecedented measurements of the mean pressure profile of clusters and the mass-observable scaling relation. These promise to enhance the accuracy of SZ and cluster-based cosmological studies.

\section*{Acknowledgements}
\small{\input{acknowledgements}}

%

%

\end{document}

%% file: acknowledgements.tex
We would like to thank the IRAM staff for their support during the observation campaigns. The NIKA2 dilution cryostat has been designed and built at the Institut N\'eel. In particular, we acknowledge the crucial contribution of the Cryogenics Group, and in particular Gregory Garde, Henri Rodenas, Jean-Paul Leggeri, Philippe Camus. This work has been partially funded by the Foundation Nanoscience Grenoble and the LabEx FOCUS ANR-11-LABX-0013. This work is supported by the French National Research Agency under the contracts "MKIDS", "NIKA" and ANR-15-CE31-0017 and in the framework of the "Investissements d’avenir” program (ANR-15-IDEX-02). This work has benefited from the support of the European Research Council Advanced Grant ORISTARS under the European Union's Seventh Framework Programme (Grant Agreement no. 291294). A. R. acknowledges financial support from the Italian Ministry of University and Research - Project Proposal CIR01$\_00010$. S. K. acknowledges support provided by the Hellenic Foundation for Research and Innovation (HFRI) under the 3rd Call for HFRI PhD Fellowships (Fellowship Number: 5357). 